\documentclass[12pt]{article}
\usepackage[ hmargin=3 cm,vmargin=3.2cm]{geometry}

 \usepackage{graphicx}
 \usepackage{amsmath}
 \usepackage{amssymb}
  \usepackage{enumerate}
\usepackage{cite}
\usepackage{slashed}

\usepackage{float}
\usepackage{tikz}
\usetikzlibrary{decorations.pathmorphing}
\usepackage[symbol]{footmisc}

\newcommand{\bfell}{\boldsymbol{\ell}}

\newcommand{\bfxi}{\boldsymbol{\xi}}

\newcommand{\C}{\mathbb{C}}

\newcommand{\bey}{\begin{eqnarray}}
\newcommand{\eey}{\end{eqnarray}}
 \begin{document} \title{Quantum field theory measurements for relativistic particles} \author {Nadia Koliopoulou\footnote{nadia.koliop@ac.upatras.gr}, \; Charis Anastopoulos\footnote{anastop@upatras.gr},  \;   and   \; Ntina Savvidou\footnote{ksavvidou@upatras.gr}\\
 {\small Laboratory of Universe Sciences, Department of Physics, University of Patras, 26500 Greece} }
\maketitle

\begin{abstract}
The formulation of a consistent measurement theory for relativistic quantum fields has become a problem of growing foundational and practical significance. Standard non-relativistic measurement models are not designed to incorporate relativistic notions
such as locality, causality and Lorentz covariance.  
While most existing work focuses on scalar fields, realistic particles possess spin, polarization, and internal degrees of freedom that introduce new conceptual and operational challenges. To this end, we  extend the  Quantum Temporal Probabilities (QTP)  framework for relativistic measurements to describe  electromagnetic, Dirac, and internally structured scalar fields. Our results include probabilities for the time-of-arrival that take spin/polarization into account, generalized photodetection formulas beyond Glauber's theory, an operational treatment of particle oscillations based on time-of-arrival measurements, together
with an analysis of the assumptions underlying the standard oscillation formula, and a measurement-theoretic analysis of relativistic qudits.
 
\end{abstract}

\section{Introduction}
In recent years, the description of quantum measurements in relativistic quantum field theory (QFT) has emerged as a focal point of foundational and practical importance---see Refs. \cite{QTP1,   OkOz, QTP3, FeVe, Bostelmanetal, FJR22, GGM22, Perche, PTM, Bednorz, PRA24}; also Refs. \cite{HK, Unruh76, Dewitt, Sorkin, PeTe} for earlier work on QFT measurements and  Refs. \cite{AHS23, MD, FeVe2} for partial reviews. Traditional quantum measurement models, rooted in non-relativistic quantum mechanics make implicit assumptions   that break down in relativistic settings, where locality, causality, and Lorentz covariance place strict constraints on what can be measured and how \cite{PeTe, AnSav22,  MD}. 

QFT makes this tension explicit: measurements follow from interactions between fields and localized detectors---the latter usually described in terms of quantum fields---and they involve   vacuum fluctuations, field correlations, and the causal structure of spacetime. As experiments and thought-experiments increasingly probe regimes where relativistic effects cannot be ignored, the need for operational frameworks that consistently incorporate relativistic notions such as
locality, causality and covariance has become increasingly apparent.  

This renewed focus has been propelled by advances at the intersection of quantum foundations, quantum information,   high-energy physics, and gravitational physics. Relativistic measurement models now provide the operational backbone for particle detection \cite{QTP3, AHS23}, entanglement in quantum fields \cite{TWPM, PPTM, AnSav25}, Unruh–Hawking phenomena \cite{Unruh76, Dewitt, AnSav12, Mou2, LBHV, AnSav20, ShCa}, and causal information transfer \cite{Sorkin, sl1,sl2, RPM, AnPl, Jubb, SPCB, Oeckl}. They clarify long-standing puzzles about state reduction, observer dependence, and the meaning of Bell locality in relativistic settings, while exposing the limits of importing non-relativistic intuitions into QFT. 

Most prior work on relativistic quantum measurements has focused on scalar field models. These provide a technically simple setting for developing detector-based measurement schemes and clarifying issues of locality and causality. While this approach has been practical, it neglects essential features of realistic quantum fields, such as spin, gauge structure, and internal degrees of freedom. 
To address physically relevant measurement scenarios, it is therefore necessary to move beyond scalar fields and incorporate these internal degrees of freedom. They introduce additional constraints and operational features that are not captured by scalar field
models.

In this paper, we analyze the measurements of relativistic particles with degrees of freedom other than translational ones (position and momentum), by considering QFT-type couplings between particles and detector. To this end, we employ the Quantum Temporal Probabilities (QTP) approach to quantum measurements\cite{QTP1, QTP2, QTP3}.

QTP was originally introduced as a general framework for describing temporally extended quantum observables \cite{AnSav06}. The basic insight underlying QTP is the clear separation between the time parameter governing Schr\"odinger evolution and the time at which a detection event occurs \cite{Sav99, Sav10}. Unlike the former, the detection time is treated as a macroscopic random variable associated with the degrees of freedom of the measuring apparatus.

The QTP approach formulates measurement theory within the decoherent-histories framework \cite{Gri, Omn1, Omn2, GeHa1, GeHa2, hartlelo}, rather than as an extension of von Neumann measurement theory \cite{vN, Busch}. A key limitation of the latter is that the time of an event is treated as a control parameter of the experiment, rather than as a random variable. In many physical setups, however, the detection time is not fixed in advance by the experimental design; it is itself a stochastic outcome. Paraphrasing Wigner \cite{Wigner}, von Neumann-type measurements are primarily suited to the question “Where is the particle now?”—fixed time, variable position—whereas QTP addresses the complementary question “When is the particle here?”—fixed detector location, variable time.

The decoherent-histories framework is appropriate because it allows one to describe measurement apparatuses that record observables not associated with a single instant of time. This is essential when the detection time is to be treated as a random variable. The histories formulation is also technically useful, because it connects the emergence of quasiclassical measurement records with the underlying dynamics of the detector. In this setting, observables correspond to suitably coarse-grained histories of the apparatus satisfying an appropriate decoherence condition.

QTP therefore treats measurement events as localized in spacetime, with their times of occurrence regarded as random variables. This reflects the structure of actual experiments, in which detectors occupy fixed spatial regions and register events at probabilistically distributed times. Each detection event is described by a set of random variables $(x,q)$, where $x$ denotes the spacetime location of the event, and $q$ denotes the remaining measured quantities, such as four-momentum, spin, or internal degrees of freedom.

In this paper, we analyze measurements pertaining to three types of fields: electromagnetic, Dirac, and scalar fields with internal degrees of freedom. 
Our central aim is to extend the QTP framework to relativistic fields carrying spin, polarization
and internal degrees of freedom, and to derive the corresponding measurement probabilities for
experimentally relevant scenarios. Specific results include  the following.
\begin{itemize}
\item We construct Positive-Operator-Valued measures (POVMs) for different types of relativistic particles. These POVMs resemble the corresponding ones for scalar particles, but they are polarization- or spin-dependent. This dependence appears at two levels: the absorption rate and the localization of the detection events. 
    
\item We analyze measurements of polarization and spin. We find that the axis of the spin/polarization measurement is determined by a two-point correlation function of the detector. 
    
\item We obtain a generalization of Glauber's photo-detection theory \cite{Glauber1, Glauber2}. In particular, we   identify  experimental regimes where additional terms might be relevant.
    
\item Our analysis clarifies the role of time-of-arrival observables in oscillation experiments and
identifies the assumptions underlying the standard oscillation formula \cite{Beuthe, Lipkin}. 

\item We define relativistic qudits in terms of internal degrees of freedom of scalar field, and we derive   a general probability formula for their measurement.
\end{itemize}


The structure of this paper is the following.   Section 2 is a review of past work, especially the QTP approach to relativistic measurements. In Section 3, we analyze measurements on photons, and in Section 4, measurements of Dirac particles. We take up composite systems in Section 5, where we study particle oscillations and relativistic qudits. In Section 6, we discuss and summarize our results.

\section{Background}

In this Section, we review the basic results of the QTP approach, and then, we use it in order to derive the time-of-arrival probability for particles with no spin or internal structure.

\subsection{The fundamental probability formula}
We analyze measurements of a quantum field $\hat{\phi}_r(x)$ on Minkowski spacetime. There is no restriction on the type of field, the index $r$ includes both spacetime and internal indices. The field operator is defined on a Hilbert space ${\cal F}$. The field interacts with  an apparatus described by a Hilbert space ${\cal H}$. 

The field and the apparatus interact through a coupling term $\int d^4x \hat{C}_a(x) \otimes \hat{J}^a(x) $ (in the interaction picture), where $\hat{C}_a(x)$ is a local composite operator for the fields $\hat{\phi}_r(x)$  and $\hat{J}^a(x)$ is a current operator on the detector's Hilbert space.

Assuming an initial state $|\psi\rangle$ for the field and an initial state $|\Omega\rangle$ for the detector,
 we derive the unnormalised probability density $P(x)$ for a detector record at a spacetime point $x$ and the measurement of an observable $\lambda$ \cite{AHS23},
\bey
P(x, \lambda) = \int d^4 y    R^{ab}(y, q) G_{ab}(x - \frac{1}{2}y, x +\frac{1}{2}y), \label{pxxx}
\eey
to leading order in the interaction. In Eq. (\ref{pxxx}),
\bey
G_{ab}(x, x')  = \langle \psi|\hat{C}_a(x) \hat{C}_b(x')|\psi\rangle
\eey
is a two-point correlation function for the field, and
\bey
R^{ab}(x, \lambda) := \langle \Omega| \hat{J}^a(x_0) \hat{\Pi}(\lambda) e^{-i\hat{P}\cdot x}\hat{J}^b(x_0)|\Omega\rangle \label{detkern}
\eey
is the {\em detection kernel}, which contains all information about the detector and the records of physical observables.

In Eq. (\ref{detkern}),  $x_0$ is a reference point in the detector's worldline,  $\hat{P}_{\mu}$ is the detector's energy momentum operator, and $\hat{\Pi}(\lambda)$ is  a   coarse-grained positive operator on the detector's Hilbert space. These formulas follow from the following assumptions.
\begin{itemize}
\item  $|\Omega\rangle$ is an eigenvector of $\hat{P}_{\mu}$, \item $\langle \Omega| \hat{J}^a(x) |\Omega \rangle = 0$, and \item the positive operator $\hat{\Pi}(\lambda)$ is dynamically stable, in the sense that $[\hat{\Pi}(\lambda), \hat{P}_{\mu}] = 0.$ 
\end{itemize}
Note that 
by definition, $R_{ab}(x, \lambda) = R^*_{ba}(-x, \lambda)$.

Eq. (\ref{pxxx}) can be derived simply by mimicking spacetime sampling with a switching function in the Hamiltonian, or in a conceptually more satisfying way, by analyzing spacetime sampling on the detector using a decoherent histories analysis of measurement records  \cite{AHS23}. The two approaches yield the same results to leading order in the field-apparatus coupling, but they differ in higher orders.

In this paper, we will consider four special cases of measurements, corresponding to different types of detection kernel:
\begin{enumerate}
    \item Ignore the observable $\lambda$---set $\hat{\Pi}(\lambda) = \hat{I}$ in the detection kernel---and focus on time-of-arrival probability distributions.
    
    \item Identify $\lambda$ with helicity or spin and consider joint measurements of time-of-arrival with helicity/spin.
    
\item Trace out over time, and consider only measurements of internal degrees of freedom (e.g., qudit measurements).

\item Identify $\lambda$ with with the energy-momentum vector $Q_{\mu}$ that is absorbed   by the detector. Then $\hat{\Pi}(Q)$ is a POVM about the detector's energy-momentum. It can be written as $\chi(\hat{P} - Q) = \prod_{\mu} \chi_\mu(\hat{P}_{\mu} - Q_{\mu})  $, where $\chi_{\mu}$ is a  sampling function for the energy-momentum coordinate $
P^\mu$, peaked around zero and normalized to unity. We will employ this POVM   for the elaboration of the time-of-arrival probabilities in composite particles, i.e., for particle oscillations. In this case, the Fourier transform $\tilde{R}^{ab}(k, Q) := \int d^4x e^{ik\cdot x}R_{ab}(x, Q)$ equals
\bey
\tilde{R}^{ab}(k, Q) = 2\pi \langle \Omega| \hat{J}^a(x_0) \delta^4(\hat{P} - k) \hat{J}^b(x_0)|\Omega\rangle \chi(k - Q). \label{ptlq}
\eey

\end{enumerate}

While the detection kernel can be defined in terms of the detector's dynamics and initial state, it can also be viewed as a semi-phenomenological quantity. In principle, it can be reconstructed from observation. Some of its components correspond to a directly measurable quantity, namely, the absorption coefficient of a detector. The remainder essentially defines the localization operator that corresponds to the intrinsic temporal spread of the measurement record.  These points are explained in Sec. 2.2.

    \subsubsection*{Poincar\'e covariance}
In QTP, we assume that  Hilbert space ${\cal F}$ of the measured system and the Hilbert space ${\cal H}$ of the apparatus carry a unitary representation of the Poincar\'e group. We also assume that the interaction-picture coupling terms  $\int d^4x \hat{C}_a(x) \otimes \hat{J}^a(x) $ transforms as a spacetime scalar. Hence, the dynamics on ${\cal H} \otimes {\cal F}$ are assumed to be Poincar\'e covariant. 

However, the POVM constructed via QTP are not so. The reason is that they depend on the initial state of the apparatus $|\Omega\rangle$, and in general, this does  not coincide  with the Poincar\'e invariant vacuum.  An actual detector involves a macroscopically large number of fermions. Hence, $|\Omega\rangle$ lies in the subspace of the fermionic fields for leptons and baryons that corresponds to macroscopically large, constant, values of the leptonic and baryonic quantum numbers. The detector defines a preferred 
reference system at which its center of momentum has zero three-momentum, and the detection POVM  

For a detector at rest in a Lorentz frame, we can assume that $|\Omega\rangle$ is approximately invariant under spacetime translations. Assuming that the same holds for the projectors $\hat{\Pi}(q)$, so does the resulting POVM. In some cases, we may even assume rotational symmetry, but in the time-of-arrival set-up we reduce to an one-dimensional problem where rotational symmetry is lost.

\subsection{Time-of-arrival measurements for scalar particles}

Next, we   review  the time-of-arrival probabilities for relativistic  scalar particles,  derived in  Ref. \cite{QTP3} via the QTP method.
 The simplest case  corresponds to a free scalar field
 $\hat{\phi}(x)$ of mass $m$,
 and field-apparatus interactions implemented through the composite operator $\hat{C}(x) = \hat{\phi}(x)$.

In a  time-of-arrival measurement, a detector is placed at a macroscopic distance $L$  from the particle source. If $L$ is much larger than the size of the detector, only particles with momentum along the axis that connects the source to the detector are recorded. Hence, the problem is  reduced to two spacetime dimensions: the scalar field $\hat{\phi}(x)$ reads,
\bey
\hat{\phi}(x) = \int \frac{dp}{\sqrt{2\pi} \sqrt{2 \epsilon_p}} \left[\hat{a}(p)e^{ipx - i \epsilon_pt} + \hat{a}^{\dagger}(p) e^{-ipx + i \epsilon_pt}\right],  
\eey
in terms of standard creation and annihilation operators.

We set the spacetime coordinate as $ x = (t, L)$. For   a field initial state with a definite number $N$ of particles, Eq. (\ref{pxxx}) yields
\begin{eqnarray}
P(t, L) = \int \frac{dpdp'}{2\pi } \frac{\rho(p,p')}{2\sqrt{\epsilon_p \epsilon_{p'}}} \; \tilde{R}\left( \frac{p+p'}{2}, \frac{\epsilon_p + \epsilon_{p'}}{2}\right) e^{i(p-p')L - i (\epsilon_p - \epsilon_{p'})t}, \label{ptx}
\end{eqnarray}
where $\epsilon_p = \sqrt{p^2+m^2}$,  $\tilde{R}(p, E)$ is the Fourier transform of $ R(x, t)$, and $\hat{\rho}(p,p') =  \langle \psi|\hat{a}^{\dagger}(p')\hat{a}(p)|\psi\rangle$ is the single-particle reduced density matrix. Note that $\tilde{R}(p, \epsilon) = 2 \pi \langle \omega |\delta(\hat{P}_0 - \epsilon) \delta(\hat{P} - p)|\omega\rangle$, where $|\omega \rangle = \hat{J}(x_0)|\Omega\rangle$. It follows that $\tilde{R}(p, \epsilon) \geq 0$ and $R(p, \epsilon) = 0 $ for $\epsilon < 0$.

In Ref. \cite{QTP3}, it is shown that the   probability distribution $P_c(t, L):= P(t, L)/P_{tot}$, conditioned upon detection, can be  expressed as
\begin{eqnarray}
P_c(t, L) = \int \frac{dpdp'}{2\pi } \tilde{\rho}(p,p')  \sqrt{v_p v_{p'}} S(p,p') e^{i(p-p')L - i (\epsilon_p - \epsilon_{p'})t}, \label{ptxb}
\end{eqnarray}
where $\tilde{\rho}(p,p')$ is an appropriately redefined initial state that accounts for the effect of conditioning on detection, and
$v_p = p/\epsilon_p$ is the particle velocity. In Eq. (\ref{ptxb}), $S(p, p')$ stands for the matrix elements $\langle p|\hat{S}|p'\rangle$ of the {\em  localization operator} $\hat{S}$, defined by
\begin{eqnarray}
 \langle p|\hat{S}|p'\rangle  := \frac{\tilde{R}\left( \frac{p+p'}{2}, \frac{\epsilon_p + \epsilon_{p'}}{2}\right)}{\sqrt{\tilde{R}(p, \epsilon_p) \tilde{R}(p', \epsilon_{p'})}}.  \label{lpp}
\end{eqnarray}
By definition, $\langle p|\hat{S}|\hat{p'}\rangle \geq 0 $ and  $S(p, p) = 1$.  The name of the operator $\hat{S}$ originates from the fact that $\hat{S}$ describes the localization of an elementary measurement event.

The conditioned probability density (\ref{ptxb}) is normalized to unity: $\int_{-\infty}^{\infty} dt P_c(t, L) = 1$ for time in the full real axis. This choice of normalization is convenient, because for states with support on positive momenta, the contribution to the integral from negative times is negligible.

For a pure initial state $|\psi\rangle$, Eq. (\ref{ptxb}) becomes
\begin{eqnarray}
P_c(t, L) = \langle \psi|\hat{U}^{\dagger}(t, L) \sqrt{|\hat{v}|}\hat{S}\sqrt{|\hat{v}|}\hat{U}(t, L)|\psi\rangle, \label{ptxc}
\end{eqnarray}
where $\hat{U}(t, L)$ is the  spacetime-translation operator
$\hat{U}(t, L) = e^{i  \hat{p} L - i \hat{H} t} $
and $\hat{v} = \hat{p}\hat{H}^{-1}$ is the velocity operator. Eq. (\ref{ptxc}) defines a positive probability distribution if and only if   $\hat{S}$ is a positive operator.  Then, the Cauchy-Schwarz inequality applies,
\begin{eqnarray}
\langle p|\hat{S}|p'\rangle \leq \sqrt{\langle p|\hat{S}|p\rangle \langle p'|\hat{S}|p'\rangle} =  1. \label{CSa}
\end{eqnarray}
Maximum localization is achieved when  Eq. (\ref{CSa}) is saturated, i.e.,  for  $\langle p|\hat{S}|p'\rangle =  1$. This occurs for $\tilde{R}(p, \epsilon) \sim \exp[-a p - b \epsilon_p]$ for $a,b > 0$. The resulting probability density was first derived by   Le\'on \cite{Leon} as a relativistic generalization of Kijowski's POVM \cite{Kij} for the time of arrival.

The maximum-localization POVM is an idealized limiting description, and not   a generic property of physical detectors. It is particularly useful, because it provides  a simple factorized form to the detection probability that is particularly useful for illustrative calculations.

\section{Photodetection}
In this section, we apply the QTP method to the quantum electromagnetic field. We construct probability densities for the measurements of the time-of-arrival in near- and far-field, and also for polarization measurements. 
\subsection{The probability formulas}
In Glauber's photodetection theory \cite{Glauber1, Glauber2}, the probability density for a photodetection event at spacetime point is
\bey
P(x) = C \langle \psi |\hat{\bf E}^{(-)}(x)\cdot \hat{\bf E}^{(+)}(x)|\psi\rangle,
\eey
where $ \hat{\bf E}^{(+)}$ ($ \hat{\bf E}^{(-)}$) stand for the positive (negative) frequency components of the quantum electromagnetic field and $C$ is a normalization constant. This expression follows from a dipole field-detector coupling and the invocation of the Rotating Wave Approximation (RWA) to drop the contribution of terms involving $\hat{\bf E}^{(-)}(x)\cdot \hat{\bf E}^{(-)}(x)$ and $\hat{\bf E}^{(+)}(x)\cdot \hat{\bf E}^{(+)}(x)$. However, the RWA is problematic at the foundational level 
 because the corresponding Hamiltonian  is unbounded from below  \cite{FoCo}. For   details about the accuracy of the RWA and related approximations, see, Refs. \cite{RWA1, FCAH10} and references therein.

In QTP, we start from Eq. (\ref{pxxx}) and assume dipole coupling between the electromagnetic field and the apparatus, that is, a coupling of the form
 $\int d^4x \hat{F}_{\mu \nu}(x) \otimes \hat{M}^{\mu \nu}(x)$. Here, $\hat{F}_{\mu \nu}$ is the electromagnetic field tensor and
$\hat{M}_{\mu \nu}$ is the electromagnetic dipole moment tensor of the detector: the $\hat{M}^{0i}$ elements correspond to the electric dipole moment, and the $\hat{M}^{ij}$ elements to the magnetic dipole moment.

 This means that the detection kernel  is of the form $R_{\mu\nu\rho\sigma}(x, q)$, with two antisymmetric pairs of indices $\mu \nu$ and $\rho \sigma$.

Here, we will consider a time-of-arrival set-up in which the field propagates from a source to a remote detector, so that the relevant field values are along one spatial dimension. Then, we can express   the field-strength tensor in terms of creation and annihilation operators as
\bey
\hat{F}^{\mu \nu}(t, x) = -i \sum_{r = 1}{2} \int \frac{dk}{\sqrt{2\pi}} \sqrt{\frac{\omega_k}{2}}( \ell^{[\mu} \epsilon^{\nu]}_r) \left[ \hat{a}_r(k) e^{ikx - i \omega_k t} - \hat{a}^{\dagger}_r(k) e^{-ikx + i \omega_k t} \right],
\eey
where $\omega_k = |k|$, $\ell^{\mu} = (1, 0, 0, 1)$, and  $\epsilon_1 = (0, 1, 0, 0)$ and $\epsilon_2 = (0,0,1,0)$ are polarization vectors.

Assuming an initial state with support only on positive momenta, Eq. (\ref{pxxx}) yields
\bey
P(t, L, \lambda) = P_0(\lambda) + Q(t, L, \lambda) + P_1(t, L, \lambda),
\eey
where
\bey
P_0(\lambda) &=& \sum_r \int \frac{dk}{4\pi} \tilde{R}_{rr}(k; \lambda), \nonumber \\
Q(t, L, \lambda) &=& - \mbox{Re} \sum_{r,r'=1}^2 \int \frac{dkdk'}{4 \pi} \sqrt{\omega_k \omega_k'} \zeta_{rr'}(k,k')  \tilde{R}_{rr'}(\frac{1}{2}(k-k'); \lambda) e^{i (k+k')L - i (\omega_k + \omega_{k'})t}, \nonumber \\
P_1(t, L, \lambda) &=&  \sum_{r,r'=1}^2 \int \frac{dkdk'}{4 \pi} \sqrt{\omega_k \omega_k'} \rho_{rr'}(k, k') \tilde{R}_{rr'}(\frac{k+k'}{2}; \lambda) e^{i (k-k')L - i (\omega_k - \omega_{k'})t}. \label{piph}
\eey
Here, we have defined the single-photon density matrix $\rho(k,k') =  \langle \psi|\hat{a}_{r'}^{\dagger}(k')\hat{a}_r(k)|\psi\rangle$, the photon-number interference function   $\zeta(k, k') = \langle \psi|\hat{a}_{r'}(k')\hat{a}_r(k)|\psi\rangle$, and
\bey
\tilde{R}_{rr'}(k; \lambda) = \ell^{\mu}\epsilon_r^{\nu} \ell^{\rho} \epsilon_{r'}^{\sigma} \tilde{R}_{\mu \nu \rho \sigma}(k, \omega_k; \lambda),
\eey
where $\tilde{R}_{\mu \nu \rho \sigma}$ is the Fourier transform of the detector kernel.

The term $P_0(\lambda)$ is  independent of the initial state. It corresponds to `vacuum noise' in our model, that is, to false-alarm detection events. It can be discarded. The term $Q(t, L,\lambda)$   is characterized by very rapid oscillations and it is typically negligible when averaging over time-scales much larger than the typical period of the EM waves. It is discarded in the RWA approximation.  

 \subsection{Far-field measurements}
Here, we analyze the time-of-arrival probabilities for far-field measurements. These correspond to the regime where the source-detector distance is much larger than the photon-wavelengths in the support of the initial state. In Sec. 3.3, we will show that in this regime 
 $Q(t, L, \lambda)$ can be ignored, so that only $P_1(t, L, \lambda)$ contributes to the total detection probability.  

We define $\tilde{R}_{rr'}(k) = \sum_\lambda \tilde{R}_{rr'}(k; \lambda)$. For fixed $k$, $R_{rr'}(k)$ is  
 a $2\times2$ positive hermitian matrix, so it can be diagonalized, and expressed as  $ \sum_{\sigma=+}^{-}  R_{\sigma}(k) \beta_r^{\sigma}(k)\bar{\beta}^{\sigma}_{r'}(k)$, where $R_{\sigma}(k)$ are the eigenvalues and $\beta^{\sigma}_r(k)$ the eigenvectors, normalized such that $\sum_{r} \beta^{\sigma}_r(k)\bar{\beta}^{\sigma'}_{r} = \delta^{\sigma \sigma'}$. 
 
We evaluate the total probability of detection $P_{tot} = \sum_{\lambda} \int_{\-\infty}^{\infty} dt P(t, L, \lambda)$. We find
$P_{tot}
= P_{tot}^{(1)} + P_{tot}^{(2)}$, where 
\bey
P_{tot}^{(\sigma)} = \sum_{rr'} \int \frac{k dk}{2} \rho_{rr'}(k,k)R_{\sigma}(k)  \beta_r^{\sigma}(k)\bar{\beta}^{\sigma}_{r'}(k). \label{ptotpol}
\eey
Each eigenvector $\beta^{\sigma}_r(k)$ defines a momentum-dependent polarization. The two eigenvectors determine an axis normal to the propagation direction for each wave number $k$. This axis corresponds to the direction of measured polarization, and it is an intrinsic feature of the apparatus,  uniquely determined by the detection kernel. Eq. (\ref{ptotpol}) implies a different absorption coefficient $\alpha_{\sigma}(k) := \frac{1}{2} k R_{\sigma}(k) $ for each polarization $\sigma$.

Next, we evaluate the conditional probability density $P_c(t, L, \sigma):= P(t, L, \sigma)/P_{tot}$. We redefine the initial state, to incorporate post-selection with  varying detection probability for different momenta, as
\bey
\tilde{\rho}_{\sigma \sigma'}(k,k') = \frac{\sqrt{\alpha_{\sigma }(k) \alpha_{\sigma '}(k')}     }{P_{tot} } \sum_{r,r'} \rho_{rr'}(k,k) \beta_r^{\sigma }\left(k \right) \bar{\beta}^{\sigma}_{r'}\left(k' \right).
\eey
Following \cite{QTP3}, it is straightforward to show that $\tilde{\rho}_{\sigma \sigma'}$ defines a normalized density matrix on $L^2({\bf R}) \otimes \C^2$. Then, we find that
\bey
P_c(t, L, \lambda) = \sum_{\sigma, \sigma'} \int \frac{dkdk'}{2\pi} \tilde{\rho}_{\sigma \sigma'}(k,k') e^{i(k-k')(L - t)} \Sigma_{\sigma \sigma'}(k, k'; \lambda) \label{pcsig}
\eey
where 
\bey
\Sigma_{\sigma \sigma'}(k, k'; \lambda) = \sum_{r,r'} \beta_r^{\sigma}(k) \bar{\beta}_{r'}^{\sigma'}(k')  \frac{\tilde{R}_{rr'}\left(\frac{1}{2}(k+k'); \lambda \right)}{\sqrt{R_{\sigma}(k) R_{\sigma'}(k')}     }
\eey
are   positive operators that incorporate both photon localization and the measurement of additional observables. 

We proceed to an analysis of special cases.

\medskip

\noindent {\bf Time-of-arrival measurements:}
First, we will consider the case of pure time-of-arrival measurements, where we sum over all other variables $\lambda$. The conditional probability $P_c(t, L) := \sum_{\lambda} P_c(t, L, \lambda)$ becomes 
\bey
P_c(t, L) = \sum_{\sigma = 1}^2   \int \frac{dkdk'}{2\pi} \tilde{\rho}_{\sigma \sigma}(k,k') S_{\sigma}(k,k') e^{i(k-k')(L - t)}, \label{eqphot}
\eey
with the localization operators $S_{\sigma}$ are defined by
\bey
S_{\sigma}(k, k') := \frac{R_{\sigma}\left(\frac{k+k'}{2}\right)}{\sqrt{R_{\sigma}(k) R_{\sigma}(k')}}. \label{slk}
\eey
Hence, the time of arrival POVM for photons is the mixture of two POVMs of the form (\ref{ptxb}) for scalar particles each corresponding to a different detected polarization. 

The maximum localization case corresponds to $S_{\lambda}(k, k') = 1$. This condition is satisfied for $\tilde{R}^{(\lambda)}(k) = C e^{-ak}$ for any $a, C \geq 0$.

In Glauber's photodetection theory, the detection kernel is simply $R_{rr'}(t, x) = \delta_{rr'} \delta(t) \delta(x)$, hence, $R^{(s)}(k)$ is a constant. Glauber's theory yields the maximum-localization POVM for $a = 0$. There is no distinction between the two polarizations, and the absorption coefficient is proportional to $k^{-1}$. 
 
 \medskip 
 
 \noindent {\bf Measuring polarization:} A   POVM for the joint measurement of polarization is obtained by identifying $\lambda$ with $\sigma$ in Eq. (\ref{pcsig}). 
  This means that we keep only the component of the detection kernel that corresponds to $\sigma$. That is, we identify 
    $\tilde{R}_{rr'}(k; \lambda)$ with the projection of the matrix $\tilde{R}_{rr'}(k)$ along the eigenvector $u^\sigma(k)$ that corresponds to polarization $\sigma$. Then, we obtain
 \bey
 P_c(t, L; \sigma) =    \int \frac{dkdk'}{2\pi} \tilde{\rho}_{\sigma \sigma}(k,k') S_{\sigma}(k,k') e^{i(k-k')(L - t)}. \label{pvm}
 \eey

 \medskip

\noindent {\bf Measurement of other observables:} If the time of detection is averaged out, the effective measured observable is time-independent, so it commutes with the Hamiltonian. Indeed, the conditional probabilities $P_c(\lambda) = \int dt P_c(t, L; \lambda)$ take the form
\bey
P_c(q) = \sum_{\lambda, \lambda'} \int dk \tilde{\rho}_{\sigma \sigma'}(k,k) \Sigma_{\sigma \sigma'}(k, k; q).
\eey
Indeed, the matrix operators $\int dk \Sigma_{\lambda \lambda'}(k, k; q)|k\rangle \langle k|$ define the most general POVM that commutes with the Hamiltonian in the single-photon Hilbert space.

\subsection{Near-field measurements}
For near-field measurements, we need to include the contribution from the term $Q(t, L; \lambda)$ in Eq. (\ref{piph}). 
For simplicity, we restrict to the time-of-arrival case (no additional variable $\lambda$) and assume a single polarization, so that the $r$-indices are suppressed.

We note that   $\int_{-\infty}^{\infty} dt Q(t, L) = 0 $. This term does not contribute to the total detection probability. Furthermore,  $Q(t, L)$ vanishes for any initial field state with a definite number of photons. 

For the general case, we use the Wigner-Weyl transform that takes the momentum matrix elements $A(k, k')$  of an operator  to a function $W_A(k, x)$, given by 
 \bey
 W_A(k, x) = \int \frac{d\xi}{2\pi} A(k+\frac{1}{2}\xi, k-\frac{1}{2}\xi) e^{i \xi x}.
 \eey
We denote the Wigner-Weyl transform of the one-particle density matrix $\hat{\rho}$ as $W_{\rho}(k, x)$ and of the photon-number interference function $\zeta(k, k')$ by $W_{\zeta}(k, x)$. By construction, $W_{\zeta}(k, x) = W_{\zeta}(k,-x)$. 

The contributions $P_1(t, L)$ and $Q(t, L)$ to the probability distribution become
\bey
P_{1}(t, L) &=& \int dk \tilde{R}(k) W_{\rho}(k, L-t) \\
Q(t, L) = &-&  \mbox{Re} \int \frac{dk dx}{\pi} R(2x) W_{\zeta}(k, x) e^{2ik(L-t)},
\eey
where $R(x)$ is the inverse Fourier transform of $\tilde{R}(k)$.

We select $\tilde{R}(k) = C e^{-ak}  \Theta(k)$ corresponding to a maximum-localization detector. Then,
\bey
P_1(t, L) &=& C \int dk e^{-ak} W(k, L-t) \label{p11}\\
Q(t, L) &=& -  Ca \mbox{Re} \int   \frac{dk dx}{\pi }  W_{\zeta}(k, x)  \frac{e^{2ik(L-t)}}{\frac{1}{4}a^2 + x^2}. \label{q11}
\eey
At the limit $a\rightarrow 0 $ that corresponds to a Glauber detector, we obtain
\bey
P_1(t, L) &=& C \int dk   W_{\rho}(k, L-t) \\
Q(t, L) &=& -C \mbox{Re} \int dk W_{\zeta}(k, 0) e^{2ik(L-t)}.
\eey
The contribution of $Q(t, L)$ is rapidly oscillatory, and it is strongly suppressed whenever the sampling $\tau$  of time measurements satisfies $k \tau >> 1$. This is easily seen by smearing $Q(t, L)$ with a Gaussian sampling function $f(t) = (2 \pi \sigma^2)^{-1/2} \exp[-\frac{t^2}{2\tau^2}]$. The smeared version of $Q(t, L)$ develops an amplitude term $e^{-\frac{1}{2}k^2 \tau^2}$ inside the integral.

On the other hand, a time-of-arrival measurement is meaningful only if $L >> \tau$. Hence,  $Q(t, L)$ can be comparable to $P_1(t, L)$ only if $W_{\zeta}$ has a significant contribution from photon frequencies with   $k L$ of order unity. In particular, this means that $Q(t, L)$ is negligible in the far-field regime. 

Note, however, that this conclusion applies only for single-event measurements. Setups that involve more that one measurements may, in principle, preserve non-RWA contributions similar to $Q(t, L)$  even in the far-field regime. 
 
 \subsubsection*{Example}
 
 We will calculate the detection probability for photons, with the EM field prepared in a coherent state $|z\rangle$. This is defined by the eigenvalue equation $\hat{a}({\bf k})|z\rangle = z({\bf k}) |z\rangle$, in terms of  square-integrable function of momentum $z({\bf k})$. For simplicity, we suppress the polarization indices. 

We straightforwardly compute
\bey
\rho(k, k')  &=& z(k) z^*(k'), \quad  \zeta(k, k')  = z(k) z(k').
\eey
Expressing $W_{\rho}$ and $W_{\zeta}$ in terms of $z(k)$, it is straightforward to show that 
\bey
P_1(t, L) &=& C |\tilde{z}(L - t)|^2, \\
Q(t, L) &=& -   C \mbox{Re} \tilde{z}(L - t)^2,
\eey
where $\tilde{z}(x)$ is the inverse Fourier transform of $z(k)$. 

For an almost monochromatic coherent state, we select $z(k) = f(k - k_0)$, where $f(k)$ is a positive-definite function that is peaked around zero. Then, $\tilde{z}(k) = \tilde{f}(L - t)e^{ik_0(L-t)}$, where $\tilde{f}$ is a positive function. Then, the total probability $P(t) = P_1(t) + Q(t)$ becomes,
\bey
P(t, L) =  C \tilde{f}^2(L - t)( 1 -   \cos[k_0(L-t)]) .
\eey
The oscillatory term  $\cos[k_0(L-t)$ is the hallmark of the breakdown of the RWA approximation.

\begin{figure}[ht]
    \centering
    \includegraphics[width=0.6\textwidth]{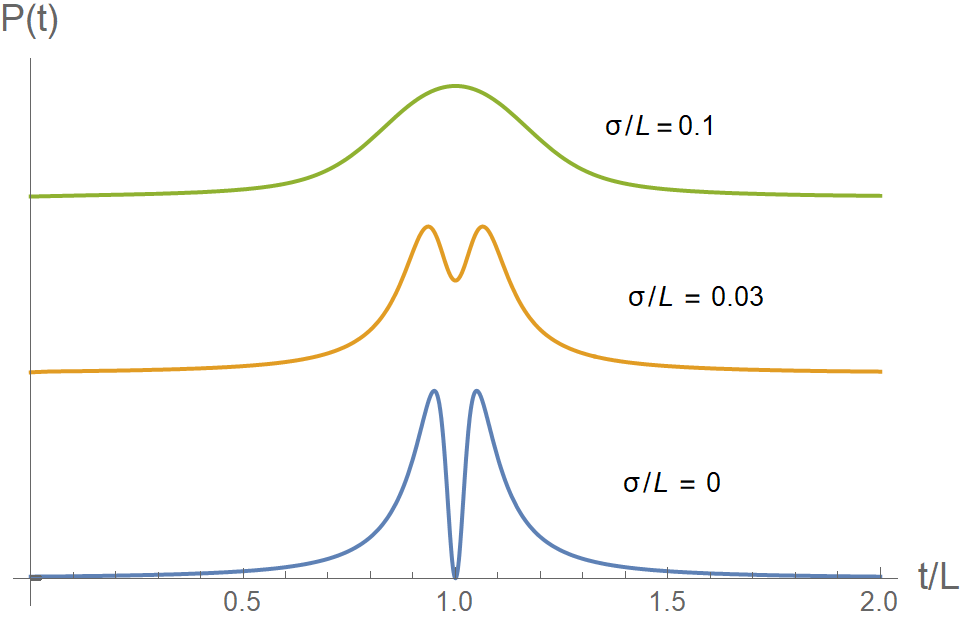} 
    \caption{Plot of $P(t, L)$ smeared with a Gaussian of width $\sigma$ in the near field regime ($k_0L = 0.5$), and for different values of $\sigma/L$. Here $f(t)$ is a Lorentzian with width $\gamma$, such that $\gamma/L = 0.05$.  }
    \label{fii}
\end{figure}

As shown in Fig. \ref{fii}, smearing with a function $f(t)$ of width $\sigma$ eventually suppresses the oscillatory term, although traces of it remain as a small drop in the probability near $t = L$.
 
\section{Dirac particles}
In this section, we analyze the measurement of relativistic particles with spin $\frac{1}{2}$, described by the Dirac field $\hat{\psi}(x)$. 
We assume that the particles are detected through electromagnetic interactions, so the composite operator   $\hat{C}_a(x)$ coincides with the (normal-ordered) current operator $ \bar{\psi}(x) \gamma^{\mu}\psi(x)$.

We will  first consider general measurements, and then specialize to time-of-arrival measurements.

\subsection{General probability formula} 

We write the Dirac field as
\bey
\psi_r(x) = \int \frac{d^3p}{(2\pi)^{3/2} \sqrt{2 \epsilon_p}} \left[\hat{c}_r({\bf p}) u_r(p)e^{i{\bf p} \cdot {\bf x} - i\epsilon_{\bf p} t} + \hat{d}_r^{\dagger}({\bf p}) \bar{v}_r(p)e^{-i{\bf p} \cdot {\bf x} +i \epsilon_{\bf p}t}  \right],
\eey
where $\hat{c}_r({\bf p})$ are annihilation operators for particles,  $\hat{d}^{\dagger}_r({\bf p})$ creation operators for anti-particles, and
\bey
u_r({\bf p}) = \frac{\slashed{p}+m}{\sqrt{2m}\sqrt{\epsilon_{\bf p}+m}} \left(\begin{array}{c}\chi_r\\0\end{array}\right),
\eey
are the associated four-spinors. Here, 
 $\chi_1 = \left(\begin{array}{c} 1\\0\end{array}\right)$ and $\chi_{-1} = \left(\begin{array}{c} 0\\1\end{array}\right)$.

For composite operators $\hat{C}_{\mu}(x) = \bar{\psi}(x) \gamma^{\mu}\psi(x)$, the detection kernel involves two spacetime indices: we write it as $R_{\mu \nu}(x; \lambda)$.

Let $|0\rangle$ be the Dirac vacuum. For an initial one-particle state $|\psi\rangle = \sum_{r} \int d^3 \psi({\bf p}) \hat{c}_r^{\dagger}({\bf p})|0\rangle$, Eq. (\ref{pxxx}) yields
\bey
P(t, {\bf L}, \lambda) = \int \frac{d^3p d^3p'}{(2\pi)^3\sqrt{2 \epsilon_{\bf p}}\sqrt{2\epsilon_{{\bf p}'}}} \psi_r({\bf p}) \psi^*_{r'}({\bf p}') Z_{rr'}({\bf p},{\bf p}'; \lambda) e^{i ({\bf p-{\bf p'})\cdot {\bf L} - i (\epsilon_{\bf p} - \epsilon_{\bf p'})t}}. \label{ferpr}
\eey
We wrote
\bey
Z_{rr'}({\bf p},{\bf p}'; \lambda) = \int d^4 y R_{\mu \nu}(y, \lambda) \left(\int \frac{d^3q}{2 \epsilon_{\bf q}}  B^{\mu \nu}_{rr'}(q, p,p') e^{i q_{\mu}y^{\mu} }\right),
\eey
where  
\bey
B^{\mu \nu}_{rr'}(q, p, p') = \frac{1}{2m}\bar{u}_r({\bf p}) \gamma^{\mu} (\slashed{q}+m)\gamma^{\nu}u_{r'}({\bf p}').
\eey
In deriving these equations, we used the identity $\sum_r u_r(p) \bar{u}_r(p) =\frac{\slashed{p} + m}{2m}$ .

Again, we can reduce the problem to one dimension, along the axis connecting source to detector, which for convenience, we take it in the $3$-axis. The detection probability becomes
\bey
P(t, L, \lambda) = \int \frac{dp dp'}{2\pi\sqrt{2 \epsilon_p}\sqrt{2\epsilon_{p'}}} \psi_r(p) \psi^*_{r'}(p') Z_{rr'}(p,p'; \lambda ) e^{i (p-p')L - i (\epsilon_p - \epsilon_{p'})t}.
\label{fer2}
\eey
Taking $p_{\mu} = (\epsilon_p, 0, 0, p)$ and similarly for $p'_{\mu}$ and $q_{\mu}$, we calculate the kernel
\bey
B^{\mu \nu}_{rr'}(q, p, p') = \delta_{rr'} \eta^{\mu \nu} B(p, p') + q_{\rho}
C^{\rho \mu \nu}_{rr'}(p, p'), 
\eey
where 
\bey
B(p,p') &=&  \frac{(\epsilon_p+m)(\epsilon_{p'}+m) - pp'}{2m \sqrt{\epsilon_p+m}\sqrt{\epsilon_{p'}+m}} \\ 
C^{\rho \mu \nu}_{rr'}(p, p') &=& \frac{1}{4m^2}\left[A^{\rho}_{rr'} \eta^{\mu \nu} + A^{\mu}_{rr'} \eta^{\rho \nu} - A^{\nu}_{rr'} \eta^{\rho \mu} - i \epsilon^{\sigma \mu \nu \rho}(p_{\lambda}+p'_{\lambda}) (\sigma^{\lambda} \bar{\sigma}^{\sigma})_{rr'}\right].
\eey
Here, we wrote 
 \bey
A^{\rho}_{rr'} = \left[(\epsilon_{p} + \epsilon_{p'}
)\delta_{rr'} + (p + p')(\sigma_3)_{rr'}, - (\epsilon_{p} + \epsilon_{p'})(\sigma_{1})_{rr'} + i(p + p')(\sigma_{2})_{rr'},
\right.
\nonumber \\
\left.
-(\epsilon_{p} + \epsilon_{p'})(\sigma_{2})_{rr'} + i(p + p')(\sigma_{1})_{rr'} , 
-(\epsilon_{p} + \epsilon_{p'})(\sigma_3)_{rr'} - (p + p')\delta_{rr'} \right].
\eey

The above expression were obtained using the identity

\begin{equation}
\gamma^\mu \gamma^\nu \gamma^\rho
=
\eta^{\mu\nu}\gamma^\rho
+
\eta^{\nu\rho}\gamma^\mu
-
\eta^{\mu\rho}\gamma^\nu
-
i \epsilon^{\sigma\mu\nu\rho}\gamma_\sigma \gamma^5 ,
\end{equation}
where 
$\gamma^5
=
i\gamma^0\gamma^1\gamma^2\gamma^3$. We use the following representation for the gamma matrices
\bey
\gamma^\mu
=
\begin{pmatrix}
0 & \sigma^\mu \\
\bar{\sigma}^\mu & 0
\end{pmatrix}, 
\eey
and the identities
\begin{align}
\sigma^0 \bar\sigma^0 = \hat{I}, \quad
\sigma^0 \bar\sigma^i = -\sigma^i, \quad
\sigma^i \bar\sigma^0 = \sigma^i, \quad
\sigma^i \bar\sigma^j = -\delta^{ij} \hat{I} - i \epsilon^{ijk}\sigma^k. \nonumber 
\end{align}
We also wrote $(\sigma^i)_{rr'} = \chi_r^\dagger \sigma^i \chi_{r'}$, since the vectors $\chi_r$ define the standard orthonormal basis on $\C^2$.


Next, we evaluate the total detection probability 
\bey
P_{tot} = \sum_{\lambda} \int_{-\infty}^{\infty} dt P(t,L, \lambda) = \int  \frac{dp}{2p} \psi_r(p) \psi_{r'}^*(p') Z_{rr'}(p),
\eey
where we wrote $Z_{rr}(p) = \sum_{\lambda} Z_{rr'}(p,p; \lambda)$.

 We diagonalize the matrix $Z_{rr'}(p)$ as $Z_{rr'}(p) = \sum_{\sigma}\zeta_{\sigma}(p) \beta^{\sigma}_r(p) \bar{\beta}_{r'}^{\sigma}(p)$, where $\zeta_{\sigma}(p)$ are the associated eigenvalues and $\beta^{\sigma}_r(p)$ the associated eigenvectors. As in the photon case, the eigenvectors $\beta^{\sigma}_r(p) $ define the direction of spin measurements in the detector, and they are fully determined by the detection kernel.

The total detection probability is
\bey P_{tot} = \sum_{\sigma}   \int dp |\psi_{\sigma}(p)|^2 \frac{\zeta_{\sigma}(p)}{2p},
\eey
where $\psi_{\sigma}(p) = \sum_r \psi_r(p) \beta_r^{\sigma}(p)$ is the wave function associated to spin $\sigma = \pm$. 
We identify the absorption coefficient for spin $\sigma$ as $\alpha_{\sigma}(p) = \frac{\zeta_{\sigma}(p)}{2p}$. The absorption coefficient is standardly defined as the fraction of particles at momentum $p$ that are absorbed by the material. 

Then, we define the post-selected states $\tilde{\psi}_{\sigma}(p) = \psi_{\sigma}(p) \sqrt{\alpha_{\sigma}(p)}/\sqrt{P_{tot}}$. The conditional probability density $P_c(t, L, \lambda) = P(t, L, \lambda)/P_{tot}$ becomes
\bey
P_c(t, L, \lambda) = \int \frac{dpdp'}{2\pi} \tilde{\psi}_{\sigma}(p) \tilde{\psi}_{\sigma'}^*(p') \sqrt{v_p v_{p'}} e^{i( p - p')L - i (\epsilon_p - \epsilon_{p'})t} S_{\sigma \sigma'}(p, p'; \lambda), \label{ptttt}
\eey
where
\bey
S_{\sigma \sigma'}(p,p';\lambda) = \sum_{rr'} \beta_r^{\sigma}(p) \bar{\beta}_{r'}^{\sigma'}(p') \frac{Z_{rr'}(p,p'; \lambda)}{\sqrt{\zeta_{\sigma}(p) \zeta_{\sigma'}(p')}}.
\eey

There is an important difference here from the case of the photon field. The spin basis that diagonalizes the total probability is not the same with the natural spin basis that 
appears when we measure spin together with the time of arrival. The latter is defined by the requirement that the localization matrix   $S_{\sigma \sigma'}(p,p';\lambda)$ is diagonal.

 \subsection{Spin measurements}
 The determination of the relativistic spin operator is a long standing problem in relativistic quantum physics, dating back to the early days of quantum mechanics\cite{Thomas, Frenkel}. The representation theory of the Poincar\'e group does not provide a unique expression, and several different candidates have been proposed---see Refs. \cite{spirev, terno} for  reviews, and references therein.. 
  
  Choosing $\lambda = \sigma$, and integrating over the time of arrival $t$, we obtain the probability for spin, conditioned on detection, as 
  \bey
  P(\sigma) = \int d^3p |\psi_{\sigma}(p)|^2 = \sum_{rr'} \int d^3p \psi_r(p) \psi^*_{r'}(p) \beta_r^{\sigma}(p) \bar{\beta}_{r'}(p).
 \eey
 The projection operators on $\C^2$ corresponding to the different values $\sigma$ of spin for a given momentum $p$ are simply $\beta_r^{\sigma}(p) \bar{\beta}_{r'}(p)$
Hence, the spin operator associated to a specific detector is simply
 \bey
 \hat{S}_{rr'} = \frac{1}{2} \int d^3p \left[ \beta^{+}_{r}(p) \bar{\beta}^+_{r'}(p) - \beta^{-}_{r}(p) \bar{\beta}^-_{r'}(p) \right]|p\rangle \langle p|.
 \eey
The explicit formula for the spin operator depends crucially on the form of the detection kernel $R_{\mu \nu}$. We will show shortly that if the detection kernel is symmetric, the spin operator is diagonal in the $rr'$ basis, i.e., the spin is measured along the direction of motion.

It is of particular interest to analyze what type of detection kernel needed to derive each spin operator that has been proposed in the literature, and whether one of the candidate operators is operationally superior to the rest. This will be undertaken in a future work.

\subsection{Time-of-arrival probabilities}
The time-of-arrival probabilities for Dirac particles are obtained by summing over $\lambda$ in Eq. (\ref{ptttt}). They take  the form
\bey
P_c(t, L) = \int \frac{dpdp'}{2\pi} \tilde{\psi}_{\sigma}(p) \tilde{\psi}_{\sigma'}^*(p') \sqrt{v_p v_{p'}} e^{i( p - p')L - i (\epsilon_p - \epsilon_{p'})t} S_{\sigma \sigma'}(p, p'), 
\eey
where $S_{\sigma \sigma'}(p,p')$ is a spin-dependent localization operator. Unlike scalar particles and photons, the time-of-arrival probabilities for Dirac particles are generically spin-dependent---at least for an EM-current type of field-apparatus coupling like the one considered here. 

The simplest case corresponds to a symmetric detection kernel $R_{\mu \nu}$. We obtain,
\bey
Z_{rr'}(p,p') =\delta_{rr'}[ a B(p,p') + \frac{b_r}{2m} [\epsilon_p + \epsilon_{p'} +  r( p + p')],
\eey
where 
\bey
a = \int \frac{dq}{2\epsilon_{q}} \eta^{\mu \nu}\tilde{R}_{\mu \nu}(q, \epsilon_q), \hspace{1cm} b_{\pm} = \frac{1}{2m} \int \frac{dq}{2\epsilon_{q}}  \eta^{\mu \nu}\tilde{R}_{\mu \nu}(q, \epsilon_q) (\epsilon_q \pm q).
\eey
All information of the detection kernel is contained in the three parameters $a, b_+$, and $b_-$.

The matrix  $Z_{rr'}(p, p')$ is already diagonal with eigenvalues 
 \bey
 \zeta_{\pm}(p, p') = a B(p, p') + \frac{b_{\pm}}{2m} [\epsilon_p + \epsilon_{p'} \pm (p+p')].
 \eey
 The absorption coefficients are 
 \bey
\alpha_{\pm} = \frac{1}{2p}   \zeta_{\pm}(p, p) =  \frac{1}{2p}[a + b_{\pm} (\epsilon_p \pm p)/m].
 \eey
 In a symmetric detection kernel,  the only available direction is the line connecting source to detector, so any spin measurement is defined with respect to this direction.
 
 Then, the time-of-arrival probability density becomes
 \bey
 P_c(t, L) = \sum_{\sigma = \pm} \int  
 \frac{dpdp'}{2\pi} \tilde{\psi}_{\sigma}(p) \tilde{\psi}_{\sigma}^*(p') \sqrt{v_p v_{p'}} e^{i( p - p')L - i (\epsilon_p - \epsilon_{p'})t} S_{\sigma}(p, p'), 
 \eey
where
 \bey
S_{\pm}(p, p') = \frac{B(p, p') + \frac{\gamma_{\pm}}{2m}  [\epsilon_p + \epsilon_{p'} \pm (p+p')]}{\sqrt{1 + \gamma (\epsilon_p \pm p)/m}\sqrt{1 + \gamma (\epsilon_{p'} \pm p')/m}},
 \eey
 with $\gamma_{\pm} = \frac{b}{ a}$. We obtain a two-parameter family of probability densities.
 
In the non-relativistic limit ($|p| < <m$), the absorption coefficient is $\alpha_{\pm}(p) = \frac{ (a+b_{\pm})}{2p}$ and the localization operator $S_{\pm}(p, p') \simeq 1$; 
they imply a  maximum localization detector for both spin degrees of freedom.
 
 In the ultra-relativistic limit ($|p| >> m$), the two spin modes behave differently at detection. The absorption coefficients become $\alpha_+ = b_+/m$ and $\alpha_- =  a/(2p)$. The localization operators become $S_+(p, p') \simeq 1$, and 
 \bey
 S_-(p, p') = \frac{p+p'}{2\sqrt{pp'}}.
 \eey
In this regime, the + modes are absorbed at a much stronger rate than the - ones.   The two helicity modes behave effectively as two distinct types of particle.

\section{ Composite relativistic particles}
In this section, we analyze measurements on composite particles, that is, particles with internal degrees of freedom.

An elementary relativistic particle is described by a unitary irreducible representation of the Poincar\'e group, and it is characterized by fixed mass $m$ and spin $s$ (and helicity $r = \pm 1 $ if $m = 0$). We write the associated Hilbert space as ${\cal H}_{m,s}$. A composite particle is described by a unitary reducible representation of the Poincar\'e group, hence, by a Hilbert space ${\cal H} = \oplus_{i} {\cal H}_{m_i, s_i}$, where $m_i$ is the mass and $s_i$ is the spin of the $i$-th internal level of the particle. It is convenient to choose the label $i = 0, 1, 2, \ldots$, so that $m_0$ corresponds to the ground state and $m_i$ is a non-decreasing sequence.
Hence, a free composite particle is characterized by a variable dispersion relation $\epsilon_{i,p} = \sqrt{m_i^2+p^2}$.  

This description is necessary when the particles are prepared in a superposition of internal states. It applies to atoms, molecules, and nuclei, but also to high energy set-ups for particle oscillations (neutral mesons and neutrinos). 

In this section, we will consider the simplest case where all internal states have spin $s = 0$. There interactions are then described by scalar fields with contributions from all internal states. We define

\bey
\hat{\phi}_V(x) = \sum_{i} \int \frac{dp}{\sqrt{2\pi} \sqrt{2\epsilon_{i,p}}} \left[V_{i} \hat{a}_{i,p}  e^{ipx - i \epsilon_{i,p}t} + V^*_{i}\hat{a}^{\dagger}_{i,p}   e^{-ipx + i \epsilon_{i,p}t}\right], \label{phif}
\eey
where $\sum_i |V_i|^2 = 1$ are unitary matrices.
Each choice of $V_{i}$ defines a different scalar field.  

In general, the matrices $V_i$ may be momentum dependent. For simplicity, we will work here the case where they are constant. 

\subsection{The particle oscillation formula}

Superpositions of masses invariably lead to oscillations, characterized by energy-dependent  frequencies and wave-lengths. In meson and neutrino oscillations, the wavelengths are macroscopic, and oscillations can be identified by measuring the total number of particles of a specific  energy, detected at a fixed distance $L$ from the source. This leads to the particle oscillation formula that connects detection probability with energy and source-detector distance---see \cite{Beuthe, Lipkin, AkSm, AkKo, Giunti} and references therein. 

Traditional derivations of the particle oscillation formula, either in QFT or in quantum mechanics, treat the particle detection as a von Neumann type measurement that takes place at a fixed moment of time. This gives rise to conceptual ambiguities that are analyzed in Refs. \cite{Lipkin, AkKo}. These   include a dependence on the initial state that could lead to alternative oscillation formulas, or different interpretations  about the observable that is actually being measured. 

A first description of particle oscillations as a time-of-arrival measurement for quantum field was presented in \cite{QTP1}. This work also used the QTP method. However, the modelling of the detectors was rather  simplified, and it did not fully constrain the general form of the probabilities . Here, we undertake a more robust   analysis.

To this end, we consider a field detector interaction with composite operator $\hat{C}_a(x) = \hat{\phi}_V(x)$. We assume a static detector at ${\bf x} = (0, 0, L)$ that records the energy $E$ and the momentum $Q$ absorbed from  the incoming particles. 

The effective field description is one-dimensional along the line that connects the source near the center of co-ordinates to detector.
The associated probability density is 
\begin{eqnarray}
P(t, L, E, Q) &=& \sum_{i,j}V_i V^*_j\int \frac{dpdp'}{2\pi } \frac{\rho_{ij}(p,p')}{2\sqrt{\epsilon_{i,p} \epsilon_{j,p'}}} \; \tilde{R}\left( \frac{p+p'}{2}, \frac{\epsilon_{i,p} + \epsilon_{j,p'}}{2}\right) 
\nonumber \\
&\times& \chi_0\left(\frac{\epsilon_{i,p}+\epsilon_{j,p}}{2} - E \right) \chi_3\left( \frac{p+p'}{2} - Q  \right)e^{i(p-p')L - i (\epsilon_{i,p} - \epsilon_{j, p'})t},\hspace{1cm} \label{ptxww}
\end{eqnarray}
where $\chi_0$ is the sample function of energy and $\chi_3$ is the sample function of momentum in the 3-direction.

For an initial state with support only on positive momenta, the time-integrated detection probability for fixed $Q$ and $E$ is
\bey
P(L, E, Q) = \sum_{ij}V_i V^*_j\int dpdp' \frac{\rho_{ij}(p,p')}{2\epsilon_{i,p} } \; \tilde{R}\left( \frac{p+p'}{2}, \epsilon_{i,p} \right)e^{i(p-p')L} \delta(\epsilon_{i,p} - \epsilon_{j, p'})\nonumber \\ 
\times \chi_0\left(\epsilon_{i,p} - E \right)\chi_3\left( \frac{p+p'}{2}  - Q  \right) 
= \int d\epsilon dq \chi_0(\epsilon -E) \chi_3(q - Q) \frac{ \tilde{R}(q, \epsilon)}{ 2q}  W(L, \epsilon, q),
\label{ptotl}
\eey
where we use the fine-grained probability density 
\bey
W(L, \epsilon,q) &=&    2 \epsilon \sum_{i,j} V_i V^*_j   \rho_{ij}(q + \frac{\Delta_{ji}}{4q}, q - \frac{\Delta_{ji}}{4q}) \nonumber \\  &\times& e^{i \frac{\Delta_{ji}L}{2q}}
\delta\left(\epsilon^2 - q^2 - \frac{m_i^2 +m_j^2}{2} - \frac{(\Delta_{ij})^2}{16q^2} \right), 
\eey
with $\Delta_{ij} = m_i^2 - m_j^2$.

Let us denote by $\sigma_E$ the spread of the energy sampling and $\sigma_Q$ the spread of momentum sampling. Assuming that the detection kernel $\tilde{R}$ to vary at scales much larger than $\sigma_E$ and $\sigma_Q$,   we can approximate
\bey
P(L, E, Q) = \frac{ \tilde{R}(Q, E)}{2Q}
\int d\epsilon dq \chi_0(\epsilon -E) \chi_1(q - Q)   W(L, \epsilon, q) \label{pinttt}
\eey
If there is only a single mass $m$, the total  probability is 
\bey
\int dE dQ P_{int}(L, E, Q) = \int dQ \frac{\tilde{R}(Q, \sqrt{E^2+m^2})}{2Q},
\eey
so the quantity $\dfrac{\tilde{R}(Q, E)}{2Q}$ coincides with the absorption coefficient of the detector. Hence, the conditioned probability density modulo detection 
\bey
P_{c}(L, E, Q) = P(L, E, Q)/ P_{tot} \nonumber 
\eey
takes the form 
$\int d\epsilon dq \chi_0(\epsilon -E) \chi_3(q - Q)   W(L, \epsilon, q) $.

When studying particles with  definite mass, the information contained in momentum measurements is identical with that contained in energy measurements. If the mass is not definite, this is no longer true. The difference is best illustrated by evaluating the 
marginal distributions of  $W(L, \epsilon, q)$. The energy marginal is 
\bey
W(L, \epsilon) = 
\int dq W(L, \epsilon, q) =  2\varepsilon \sum_{i,j} V_i V_j^* \rho_{ij}(p_i(\epsilon),p_j(\epsilon)) \nonumber \\
\times \left( \frac{e^{-i\frac{\Delta_{ij}L}{p_i(\epsilon)+p_j(\epsilon)}}}{p_i(\epsilon)^{-1}+p_j(\epsilon)^{-1}} + \frac{e^{-i\frac{\Delta_{ij}L}{|p_i(\epsilon)-p_j(\epsilon)|}}}{|p_i(\epsilon)^{-1}-p_j(\epsilon)^{-1}|}\right),
\eey
 where we wrote $p_i(\epsilon) = \sqrt{\epsilon^2 - m_i^2}$. 
The momentum marginal is
\bey
W(L, q) =  
\int d\epsilon  W(L, \epsilon, q) = \sum_{i,j} V_i V_j^*
 \rho_{ij}\!\left(q+\frac{\Delta_{ij}}{4q}, q - \frac{\Delta_{ij}}{4q}\right)
e^{-i\frac{\Delta_{ij} L}{2q}}. \label{wlq}
\eey

The detection probability for  momentum measurements oscillates with characteristic wave number $K(q) = \frac{\Delta_{ij}}{2q}$ at momentum $q$. In contrast, the detection probability for energy measurements has two characteristic oscillation wave numbers: $K_1(\epsilon) = \frac{\Delta_{ij} }{p_i(\epsilon) + p_j(\epsilon)}$ and $K_2(\epsilon) = 
\frac{\Delta_{ij} }{|p_i(\epsilon)- p_j(\epsilon)|}$ with different amplitudes.

Consider the regime where the kinetic energy of the particles is much larger than the mass differences. In this regime, $|p_i(\epsilon)/q - 1| < \sqrt{|\Delta_{ij}|}/q << 1$, and we find that 
$K_1(\epsilon) \simeq \frac{\Delta_{ij} }{p_i(\epsilon)}$, and $K_2(\epsilon) \simeq  2 p_i(\epsilon)$. The wave-number $K_1(\epsilon)$ then  coincides with that of momentum measurements up to an error of order $\sqrt{|\Delta_{ij}|}/q$, and $K_2(\epsilon) >> K_1(\epsilon)$ is much larger corresponding to very fast oscillations with a large amplitude. 

The $K_2(\epsilon)$ oscillations are suppressed when the probability density is smeared with $\chi_0(\epsilon -E)$ to obtain the observable probabilities. For a Gaussian smearing function with width $\sigma_E$, they are  suppressed by a factor $\exp[- \sigma_E^2L^2/ v_E^2]$, which becomes very small for sufficiently large distances $L$. 

We conclude that the standard oscillation formula is obtained independently of the type of measurements we perform, as long as the kinetic energy of the particles is much larger than the mass differences and the detector is in the far field regime: $\sigma_E L/v_E >> 1$. Outside this regime,  the oscillation pattern depends crucially on whether we record energy or momentum in the detector.

\subsection{Dependence on the initial state}

The oscillation frequencies that we derived are purely dynamical, they contain no input from the initial state. It is conceivable that the measured probabilities involve additional contributions from phase factors in the  initial state.  
As suggested by Schwinger \cite{Schwinger1, Schwinger2}, the action of external sources on the vacuum provides a mathematically well-defined, and operationally meaningful procedure for generating initial states in QFT.
 
Therefore, we assume that the initial state is generated by  an interaction Hamiltonian 
\bey
\hat{H}_{int} = \int dt dx J(x, t) \hat{\phi}_{U}(x,t), 
\eey
involving a field $\hat{\phi}_{U}$ and an external source  $J(x, t)$. We assume that the latter is localized near $x = 0$ and is non-zero for    times  $t  < 0$. For an initial field vacuum $|0\rangle $, the state of the field at $t = 0$ is  $|\Psi_0\rangle = {\cal T} \exp[-i \int dt dx J(x, t) \hat{\phi}_U(x, t)]|0\rangle$, where ${\cal T}$ denotes time ordering. Using Magnus identity \cite{Magnus}, we find that 
\bey
 |\Psi_0\rangle = e^{i\theta} \exp[-i \int dt dx J(x, t) \hat{\phi}_U(x, t)]|0\rangle,
\eey
for some phase $\theta$.
Hence, the single-particle density matrix $\rho_{ij}(p,p') = \langle \Psi_0|\hat{a}^{\dagger}_j(p') \hat{a}_i(p)|\Psi_0\rangle $ is of the form $\psi_i(p) \psi^*_j(p')$, with
\bey
\psi_i(p) = (2\epsilon_{i,p})^{-1/2} U_i \tilde{J}(p, -\epsilon_{i,p}), \label{currp}
\eey
where $\tilde{J}$ is the Fourier transform of $J$.

 It follows that
\bey
W(L, \epsilon, q) =     \sum_{i,j} V^*_i U_i V_j U^*_j   \tilde{J}(q + \frac{\Delta_{ji}}{4q}, - \epsilon) \tilde{J}^*(q - \frac{\Delta_{ji}}{4q}, -\epsilon)   e^{i \frac{\Delta_{ji}L}{2q}}
\nonumber \\
\times
\delta\left(\epsilon^2 - q^2 - \frac{m_i^2 +m_j^2}{2} - \frac{(\Delta_{ij})^2}{16q^2} \right),  \label{wlq2}
\eey
We see that a  generic localized source term does not typically generate additional phases in Eq. (\ref{wlq2}). For any source that is concentrated as scales much smaller than the oscillation wavelength, we can approximate $\tilde{J}(q + \frac{\Delta_{ji}}{4q}, - \epsilon) \simeq \tilde{J}(q, - \epsilon)$, to obtain
\bey
W(L, \epsilon, q) =  |J(q, -\epsilon)|^2   \sum_{i,j} V^*_i U_i V_j U^*_j      e^{i \frac{\Delta_{ji}L}{2q}}
\delta\left(\epsilon^2 - q^2 - \frac{m_i^2 +m_j^2}{2} - \frac{(\Delta_{ij})^2}{16q^2} \right), \label{wleq}
\eey
 which demonstrates that there is no contribution to the oscillations from the initial state.

\subsection{Time of arrival probabilities}

To obtain time-of-arrival probabilities, we integrate over $E$ and $Q$ in Eq. (\ref{ptxww}). If we assume that the detection kernel varies slowly at the scale of $\Delta_{ij}$, we can straightforwardly apply the methods of previous sections, to derive the time of arrival probability density, conditioned on detection.

Integrating over $E$ and $Q$, we can define the post-selected time-of-arrival probability as 
 \begin{eqnarray}
P_c(t, L) = \sum_{i,j}V_i V^*_j\int \frac{dpdp'}{2\pi }  \rho_{ij}(p,p') \sqrt{v_{ip} v_{jp'}} S(p,p')  e^{i(p-p')L - i (\epsilon_{i,p} - \epsilon_{j, p'})t}, \label{ptxww2}
\end{eqnarray}
in terms of a single localization operator $S(p, p')$.

 For maximal localization, and for a pure initial state $\tilde{\psi}_i(p)$, the probability density takes the simple form
 \bey
P_c(t, L) = \sum_{ij} {\cal A}_i(t, L) {\cal A}^*_j(t, L)
\eey
where 
\bey
{\cal A}_i(t, L) = V_i \int \frac{dp}{\sqrt{2\pi}}  \tilde{\psi}_i(p) \sqrt{v_{ip}} e^{ipL - i \epsilon_{i, p}t}.
\eey

An almost monochromatic initial state of the form (\ref{currp}) with momentum $p_0$ can be expressed as $\tilde{\psi}_i(p) = U^*_i \tilde{\psi}(p- p_0)$, where $\tilde{\psi}(p)$ is a real-valued function  strongly peaked around $0$. Then, ignoring wave function dispersion, we can approximate
\bey
{\cal A}_i(t, L) = V_iU^*_i \sqrt{v_{i,p_0}} e^{ip_0L  -i\epsilon_{i, p_0}t}  \psi(L - v_{i,p_0}t), \label{mmma}
\eey
where $\psi(x)$ is the inverse Fourier transform of $\tilde{\psi}(p)$. Hence, 
\bey
P_c(t,L) = \sum_{ij}\sum_{i,j} V^*_i U_i V_j U^*_j \sqrt{v_{i,p_0} v_{j,p_0}} e^{-i (\epsilon_{i, p_0} - \epsilon_{j, p_0})t}  \psi(L - v_{i,p_0}t) \psi(L - v_{j,p_0}t).\label{pcss0}
\eey
The phase difference in the exponent is of the same order as the difference in the arguments of $\psi(L - v_{i,p_0}t)$ giving as the peaks of the wave-functions. For $|\Delta_{ij}|<< p_0$, it is convenient to define $\epsilon_0 = \sqrt{p_0^2 + M^2}$, where $M$ is the median of the masses $m_i$. We also write $v_0 = p_0/\epsilon_0$, and express the time-of-arrival probabilities in terms of the variable $s = v_0 t - L$. We obtain
\bey
P_c(s, L) = \sum_{ij}\sum_{i,j} V^*_i U_i V_j U^*_j e^{i\frac{\Delta_{ij}L}{2p_0}}\left[e^{i\frac{\Delta_{ij}s}{2p_0}}  \psi(s - L \frac{\delta_i}{2 p_0^2}) \psi(s - L \frac{\delta_j}{2 p_0^2}) \right], \label{pcss}
\eey
where $\delta_j = m_i^2 - M^2$. 

It is important to emphasize that the oscillations in the probability density (\ref{pcss0}) are very different from the ones of the probability density (\ref{ptotl}). The two probability densities refer to different observables: time of arrival versus energy and momentum. Since these observables do not commute, they are incompatible, and any transformation that purports to derive one from the other is not justified by the rules of quantum theory. 

We see that the two probability densities have very different properties. The probability density (\ref{pcss}) strongly depends on the shape of the initial state $\psi(x)$, while in the probability density (\ref{ptotl}) the initial state only determines the relative weight of different energies in the initial wave packet. 
Eq. (\ref{pcss}) implies that there is no oscillatory behavior in the time of arrival, if $\frac{L\Delta_{ij}}{p_0^2}$ is much larger than the width $\sigma$ of $\psi(x)$, i.e., for $L > L_{coh} = \sigma p_0^2/\Delta_{ij}$. No such condition is relevant to the probability density (\ref{ptotl}). Particle oscillations---as measured by the detection probability---do not decohere at very long distances. 

This is an important point, because it resolves a commonly encountered ambiguity in the derivation of the oscillation formula. One encounters amplitudes similar to Eq. (\ref{mmma}), when analyzing the propagation of the wave-packet. One then constructs a detection probability by choosing a specific time $t$ that corresponds to the peak of the wave-packet. This procedure is not justified fundamentally, and, certainly, it is strongly state-dependent. Our analysis does not require such {\em ad hoc} assumptions, as it works with different POVMs for each case.

\subsection{Relativistic qudits}
The definition of relativistic qudits is an important problem in relativistic quantum information. A relativistic system must carry unitary representations of the Poincar\'e group, which are all infinite-dimensional, while qudits are  defined in the finite dimensional Hilbert space $\C^d$.

We can describe relativistic qudits as particles with $d$ internal degrees of freedom, as described in this section. We operate on the qudits through scalar fields  $\hat{\phi}_V(x)$ and interactions of the form $\int d^4 x \hat{\phi}_V(x) \hat{J}(x)$ with external currents $\hat{J}(x)$. The idea is that we can access different qudit initial states by controlling the interaction channels through which they are generated and detected.

Hence, for particles generated through interactions with fields $\hat{\phi}_V(x)$ and detected through interactions with fields $\hat{\phi}_U(x)$ at distance $L$ from the source, the total detection probability is constructed by integrating Eq. (\ref{wlq2}) with respect to $\epsilon$ and $q$. We obtain
\bey
P_{tot}  =  \int dq \sum_{i,j} V^*_i U_i V_j U^*_j  A_{ij}(q), \label{ptotd}
\eey
where 
\bey
A_{ij}(q) = \frac{1}{2 \sqrt{\epsilon_{ij}(q)}}  \tilde{J}(q + \frac{\Delta_{ji}}{4q}, - \epsilon_{ij}(q)) \tilde{J}^*(q - \frac{\Delta_{ji}}{4q}, -\epsilon_{ij}(q))   e^{i \frac{\Delta_{ji}L}{2q}},
\eey
with $\epsilon_{ij}(q) = \sqrt{ q^2 + \frac{m_i^2 +m_j^2}{2} + \frac{(\Delta_{ij})^2}{16q^2}}$. For $\Delta_{ij} << q^2$, we can substitute $\epsilon_{ij}(q)$ with $\epsilon_q =  \sqrt{q^2 + M^2}$, so that
\bey
A_{ij}(q) = \frac{1}{2 \epsilon_q} |\tilde{J}(q, -\epsilon_q)|^2 e^{i \frac{\Delta_{ji}L}{2q}},
\eey
and the dependence on $i$ and $j$ is carried solely by the oscillation phase $e^{i \frac{\Delta_{ji}L}{2q}}$. 

We rewrite Eq. (\ref{ptotd}) as follows. Let $\hat{M}$ be the mass operator on $\C^d$ with eigenstates $|i\rangle$, so that $\hat{M}|i\rangle = m_i |i\rangle$. The probability of obtaining a final qudit state $|V\rangle = \sum V_i|i\rangle$ given an initial state $|U\rangle = \sum_i U_i|i\rangle$ is given by
\bey
P_{tot} = \int dq f(q) |\langle V|e^{-i\hat{M}^2L/q}|U\rangle|^2, \label{ptotd44}
\eey
where $f(q) = \frac{|\tilde{J}(q, -\epsilon_q)|^2}{2 \epsilon_q}$ gives the momentum distribution of the initial state.

 Eq. (\ref{ptotd44}) is the fundamental equation for probabilities in a relativistic qudit. It relates purely operational quantities: the interaction Hamiltonian generating the qudit as expressed in $f(q)$ and the vector $|U\rangle$, and the interaction Hamiltonian guiding detection as expressed in the vector $|V\rangle$.

\section{Conclusions}
 We have described our motivation and main results in the introduction.  
Here, we want to focus on the applications  of our results.
 
 First, our treatment of photodetection and the characterization of the deviations from Glauber's theory is of direct relevance  to long-baseline quantum experiments in space \cite{Rideout, DSQL}, where such deviations are, in principle, detectable. The QTP formalism applies directly to such set-ups, and, crucially, it can take into account the effect of the detectors' motion into the probabilities.

 Second, our analysis of time-of-arrival measurements of Dirac particles uncovered a strong spin-dependence on the time-of-arrival probabilities in the relativistic regime. It is important to check whether this phenomenon is directly measurable, and to understand its implications. 

 Third, our Dirac particle analysis provides a basis for an operational solution to a long-standing problem of relativistic quantum theory: the correct definition of a spin operator. We will undertake this analysis in future work, paying particular attention to its behavior under changes of Lorentz frames.

 Fourth, our description of systems with internal degrees of freedom provides a concrete operational definition of relativistic qudits, and the corresponding probability formula for qudit measurements. Further work is needed, to identify the behavior of such qudits under changes of Lorentz frame. The overall formalism is particularly appropriate for the analysis of free-fall for composite quantum particles, especially regarding the generalization of the equivalence principle in the quantum domain \cite{ZyBru, Zych, AnHu18}.

 Finally, we note that our methodology may be transferrable to other approaches to QFT measurements, thus, enhancing their applicability to realistic experiments.

 \section*{Acknowledgements}
 C.A. and K.S acknowledge support by   grant  JSF-19-07-0001 from the Julian Schwinger Foundation for the early stages of this project. K.S. acknowledges support from the Andreas Mentzelopoulos Foundation through the  Branded Academic Position on ``Quantum Science and Technology" at the University of Patras.


\begin{thebibliography}{}



\small

 
 \bibitem{QTP1}	C. Anastopoulos and N. Savvidou, {\em Time-of-Arrival Probabilities for General Particle Detectors}, Phys. Rev. A86, 012111 (2012).





 \bibitem{OkOz} K. Okamura and M. Ozawa, {\em Measurement Theory in Local Quantum Physics}, J. Math. Phys. 57, 015209
(2015).


 
\bibitem{QTP3}	C. Anastopoulos and N. Savvidou, {\em Time of Arrival and Localization of Relativistic Particles}, J. Math. Phys. 60, 0323301 (2019).

\bibitem{FeVe} C. J. Fewster and R. Verch, {\em Quantum Fields and Local Measurements}, Comm. Math. Phys. 378, 851 (2020).

\bibitem{Bostelmanetal} H. Bostelmann, C. J. Fewster, and M. Ruep {\em Impossible Measurements Require Impossible Apparatus}, Phys. Rev. D 103, 025017 (2021).


\bibitem{FJR22}C. J. Fewster, I. Jubb, and M. H. Ruep, {\em Asymptotic Measurement Schemes for Every Observable of a Quantum Field Theory},  Ann. Henri Poincaré 24, 1137 (2023).



\bibitem{GGM22}  J. Polo-Gómez, L. J. Garay, L. J. and E. Martín-Martínez,  {\em A Detector-Based Measurement Theory for Quantum Field Theory},  	Phys. Rev. D 105, 065003 (2022).

\bibitem{Perche} T. R. Perche, {\em Localized Nonrelativistic Quantum Systems in Curved Spacetimes: A General Characterization of Particle Detector Models},
Phys. Rev. D106, 025018 (2022).

\bibitem{PTM} T. R. Perche, J. Polo-Gómez, B. de S. L. Torres, and E. Martín-Martínez, {\em Particle Detectors from Localized Quantum Field Theories},  	Phys. Rev. D109, 045013 (2024).

 

\bibitem{Bednorz} A. Bednorz, {\em General Quantum Measurements in Relativistic Quantum Field Theory}, Phys. Rev. D108, 056020 (2023).

\bibitem{PRA24} M. Papageorgiou, J. de Ramón, and C. Anastopoulos. {\em Particle-Field Duality in QFT Measurements}, Phys. Rev. D109, 065024 (2024).


\bibitem{HK} K. E. Hellwig and K. Kraus, {\em Operations and Measurements. II}, Comm. Math. Phys. 16, 142 (1970). 


\bibitem{Unruh76} W. G. Unruh, {\em Notes on Black Hole Evaporation}, Phys. Rev. D14, 870 (1976).

\bibitem{Dewitt}B. S. DeWitt, {\em Quantum Gravity: the New Synthesis} in
``General Relativity: An Einstein Centenary Survey", eds. S. W. Hawking
and W. Israel (Cambridge University Press, Cambridge, 1979).




\bibitem{Sorkin} R. Sorkin, {\em Impossible Measurements on Quantum Fields}, in ``Directions in General Relativity”, eds. B. L. Hu and T. A. Jacobson (Cambridge University Press, Cambridge  1993).



\bibitem{PeTe} A. Peres and D. Terno,   {\em Quantum Information and Relativity Theory}, Rev. Mod. Phys. 76, 93 (2004).

 

\bibitem{AHS23} 
 C. Anastopoulos, B. L. Hu, and K. Savvidou,{\em Quantum Field Theory Based Quantum Information: Measurements and Correlations}, Ann. Phys. 450,  169239 (2023). 
 



\bibitem{MD} M. Papageorgiou and D. Fraser, {\em Eliminating the `Impossible': Recent Progress on Local Measurement Theory for Quantum Field Theory}, Found. Phys.  54 (2023).


\bibitem{FeVe2} C. J. Fewster and R. Verch, {\em Measurement in Quantum Field Theory}, Encyclopedia of Mathematical Physics, 
5, 335 (2025).
 

\bibitem{AnSav22} C. Anastopoulos and N. Savvidou,	{\em Quantum Information in Relativity: The Challenge of QFT Measurements},  Entropy 24, 4 (2022).
 


\bibitem{TWPM}    B.  Torres, K. Wurtz, J. Polo-Gómez, and  E. Martín-Martínez, {\em Entanglement Structure of Quantum Fields through Local Probes}, JHEP 2023, 58 (2023).

\bibitem{PPTM} T. R. Perche, J. Polo-Gómez,   B.  Torres, and  E. Martín-Martínez, {\em Fully Relativistic Entanglement Harvesting}, Phys. Rev. D 109, 045018 (2024).

\bibitem{AnSav25}C. Anastopoulos  and N. Savvidou, {\em 	Relativistic Quantum Information Measures from Unequal-Time QFT Correlation Functions},   Ann. Phys. 481, 170146 (2025). 



\bibitem{AnSav12}	C. Anastopoulos and N. Savvidou, {\em Coherences of Accelerated Detectors and the Local Character of the Unruh Effect}, J. Math. Phys. 53, 012107 (2012).



\bibitem{Mou2} D. Moustos, {\em Asymptotic States of Accelerated Detectors and Universality of the Unruh Effect}, Phys. Rev. D98, 065006 (2018).


\bibitem{LBHV} C. A. U. Lima, F. Brito, J. A. Hoyos and D. A. T.  Vanzella, {\em Probing the Unruh Effect with an Accelerated Extended System}, Nature Comm. 10, 3030 (2019).



\bibitem{AnSav20}  C. Anastopoulos and N. Savvidou, {\em Multi-Time Measurements in Hawking Radiation: Information at Higher-Order Correlations}, Class. Quant. Grav. 37, 025015 (2019).
    
    \bibitem{ShCa} C. J. Shallue and S. M. Carroll, {\em What Hawking Radiation Looks Like as you Fall into a Black Hole}, Phys. Rev. D 112, 085013 (2025).
    
    
\bibitem{sl1} M. Cliche and A.Kempf, {\em The Relativistic Quantum Channel of Communication Through Field Quanta}, Phys. Rev. A 81, 012330 (2010).

 \bibitem{sl2}  E. Martin-Martinez, {\em Causality Issues of Particle Detector Models in QFT and Quantum Optics}, Phys. Rev. D 92, 104019 (2015).
    
    \bibitem{RPM} J. de Ramon, M. Papageorgiou and E. Martin-Martinez, {\em Relativistic Causality in Particle
Detector Models: Faster-than-light Signalling and ”Impossible measurements”}, Phys. Rev. D103, 085002 (2021).
    
    \bibitem{AnPl} C. Anastopoulos and M. E. Plakitsi, {\em 	Relativistic Time-of-Arrival Measurements: Predictions, Post-Selection and Causality Problems},   Foundations 3, 724 (2023). 
    
    \bibitem{Jubb} I. Jubb. {\em Causal State Updates in Real Scalar Quantum Field Theory}, Phys. Rev. D105, 025003 (2022).
    
    \bibitem{SPCB}R. Simmons, M. Papageorgiou, M. Christodoulou, and  Č. Brukner, {\em Factorisation Conditions and Causality for Local Measurements in QFT},  	arXiv:2511.21644.
        
        
        \bibitem{Oeckl} R. Oeckl, {\em Causal Measurement in Quantum Field Theory: spacetime},  	arXiv:2511.06566.
        
    

\bibitem{QTP2}	C. Anastopoulos and N. Savvidou, {\em Time-of-Arrival Correlations}, Phys. Rev. A95, 032105 (2017).



 \bibitem{AnSav06} C. Anastopoulos and N. Savvidou, {\em Time-of-Arrival Probabilities and Quantum Measurements},  J. Math. Phys. 47, 122106 (2006).




\bibitem{Sav99} K. Savvidou,  {\em The Action Operator for Continuous-time Histories}  J. Math. Phys. 40, 5657 (1999); {\em Continuous Time in Consistent Histories}, gr-qc/9912076.

\bibitem{Sav10} N. Savvidou, {\em Space-time Symmetries in  Histories Canonical Gravity}, in ``Approaches to Quantum Gravity", edited by D. Oriti (Cambridge University Press, Cambridge 2009).


 

\bibitem{Gri} R. B. Griffiths, {\em 	Consistent Quantum Theory} (Cambridge University Press, Cambridge 2003).
    
\bibitem{Omn1} R. Omn\'es, {\em The Interpretation of Quantum Mechanics}, (Princeton University Press, Princeton 1994).

\bibitem{Omn2}  R. Omn\'es,	{\em Understanding Quantum Mechanics} (Princeton University Press, Princeton 1999).




\bibitem{GeHa1} M. Gell-Mann and J.  B.  Hartle,  {\em Quantum Mechanics in
the Light of Quantum Cosmology}, in ``Complexity, Entropy, and the Physics of Information",   ed. by W. Zurek,
(Addison Wesley, Reading 1990);

\bibitem{GeHa2} M. Gell-Mann and J.  B.  Hartle, {\em Classical Equations
for Quantum Systems},  Phys. Rev.   D47, 3345 (1993).



\bibitem{hartlelo} J.B. Hartle,
{\em Spacetime Quantum Mechanics and the
Quantum Mechanics of Spacetime}
in ``Gravitation and Quantizations",  
Proceedings of the 1992 Les Houches Summer School, ed. by B. Julia and J. Zinn-
Justin,  Les  Houches  Summer  School  Proceedings,
Vol. LVII, (North Holland, Amsterdam, 1995); [gr-qc/9304006].

 \bibitem{Wigner} E.P. Wigner, {\em On the Time-Energy Uncertainty Relation},  in ``Aspects of Quantum
Theory", ed. by A. Salam, E.P. Wigner (Cambridge University Press,  1972).


\bibitem{vN} J. von Neumann, {\em Mathematical Foundations of Quantum Mechanics} (Princeton University Press, Princeton, 1955).

\bibitem{Busch} P. Busch, P. J. Lahti and P. Mittelstaedt, {\em The Quantum Theory of Measurement},  Lecture Notes in Physics Monographs, volume 2 (1996).


\bibitem{Glauber1} 	R. J. Glauber, {\em The Quantum Theory of Optical Coherence}, Phys. Rev. 130, 2529 (1963).

 \bibitem{Glauber2} 	R. J. Glauber, {\em Coherent and Incoherent States of the Radiation Field}, Phys. Rev. 131, 2766 (1963).
 
 
  

\bibitem{FoCo}  G. W. Ford and R. F. O'Connell, {\em The Rotating Wave Approximation (RWA) of Quantum Optics: serious defect}, Physica A243, 377 (1997).


 

\bibitem{RWA1} G. S. Agarwal, {\em Rotating-Wave Approximation and Spontaneous Emission}, Phys. Rev. A4, 1778 (1971);


\bibitem{FCAH10} C. Fleming, N. I. Cummings, C. Anastopoulos and B. L. Hu,  {\em The Rotating-Wave Approximation: Consistency and Applicability from an Open Quantum System Analysis}, J. Phys. A: Math. Theor. 43, 405304 (2010).



 
 
 \bibitem{Beuthe} M. Beuthe, {\em Oscillations of Neutrinos and Mesons in Quantum Field Theory}, Phys. Rep. 375, 105 (2003).

\bibitem{Lipkin} H. J. Lipkin, {\em Quantum Theory of Neutrino Oscillations for Pedestrians: Simple Answers to Confusing Questions}, Phys. Lett. B642, 366 (2006).
 
 \bibitem{Leon} J. Le\'on, {\em Time-of-Arrival Formalism for the Relativistic Particle}, J. Phys A: Math. Gen. 30, 4791 (1997).



   \bibitem{Kij} J. Kijowski,  {\em On the Time Operator in Quantum Mechanics and the Heisenberg Uncertainty Relation for Energy and Time}, Rep. Math. Phys. 6, 361 (1974).

\bibitem{Thomas} L. H. Thomas, {\em The Kinematics of an Electron with an Axis}, Phil. Mag. 3, 1 (1927).

\bibitem{Frenkel} J. Frenkel, {\em Die Elektrodynamik des Rotierenden Elektrons}, Z. Phys. 37, 243 (1926).

\bibitem{spirev}H. Bauke, S. Ahrens, C. H.  Keitel and R.  Grobe, {\em What is the Relativistic Spin Operator?}, New J. Phys. 16, 043012 (2014).


 \bibitem{terno} D. Terno, {\em Two Roles of Relativistic Spin Operators}, Phys. Rev. A 67, 014102 (2003).
 

\bibitem{AkKo} E. K. Akhmedov and J. Kopp, {\em Neutrino Oscillations: Quantum Mechanics vs. Quantum Field Theory},  	JHEP 1004:008 (2010).

\bibitem{AkSm}E. K.  Akhmedov and E. Y. Smirnov, {\em Paradoxes of Neutrino Oscillations}, Phys. Atom. Nucl.72. 1363 (2009).

\bibitem{Giunti} C. Giunti and  C. W. Kim, {\em Quantum Mechanics of Neutrino Oscillations},  (Found. Phys. Lett. 14, 213 (2001).


\bibitem{Schwinger1} J. Schwinger, {\em Particles and Sources}, Phys. Rev. 152, 1219 (1966).

\bibitem{Schwinger2}J. Schwinger, {\em Particles, Sources, and Fields}, (Reading, Mass., Perseus Books 1998).

\bibitem{Magnus} W. Magnus, {\em On the Exponential Solution of Differential Equations for a
Linear Operator}, Comm. Pure Appl. Math.  7, 649 (1954).

\bibitem{Rideout}	D. Rideout et al, {\em Fundamental Quantum Optics Experiments Conceivable with Satellites -- Reaching Relativistic Distances and Velocities}, Class. Quantum Grav. 29, 224011 (2012).

\bibitem{DSQL}M. Mohageg et al, {\em The Deep Space Quantum Link: Prospective Fundamental
Physics Experiments using Long-Baseline Quantum Optics},  EPJ Quantum Technol. 9, 25 (2022).

\bibitem{ZyBru}
M. Zych and C. Brukner, {\em Quantum Formulation of the Einstein Equivalence Principle},  Nature Phys 14, 1027  (2018). 

\bibitem{Zych} M. Zych, {\em Quantum Systems under Gravitational Time Dilation}, (Springer, 2017).

\bibitem{AnHu18}C. Anastopoulos and B. L. Hu, {\em Equivalence Principle for Quantum Systems: Dephasing and Phase Shift of Free-Falling
Particles},  Class. Quant. Grav. 35, 035011 (2018).
 
\end{thebibliography}
\end{document}